\documentclass[11pt,twoside]{article} 
\usepackage{times,fancyhdr}
\usepackage{cite}
\usepackage{graphicx}
\usepackage{amsmath} 

\date{}

\title{Complex Network view of performance and risks on Airport Networks} 
\author{Ganesh Bagler\\ National Centre for Biological Sciences, Tata Institute of Fundamental Research,\\ Bangalore, India 560065} 

\begin{document} 

\pagestyle{fancy}
\fancyhead{} 
\fancyhead[EC]{Ganesh Bagler}
\fancyhead[EL,OR]{\thepage}
\fancyhead[OC]{Complex Network view of performance and risks on Airport Networks}
\fancyfoot{} 
\renewcommand\headrulewidth{0.5pt}
\addtolength{\headheight}{2pt} 

\maketitle 

\section*{Abstract}
Air transportation has been becoming a major part of transportation infrastructure worldwide. Hence the study of the Airports Networks, the backbone of air transportation, is becoming increasingly important. In complex systems domain, airport networks are modeled as graphs (networks) comprising of airports (vertices or nodes) that are linked by flight connectivities among the airports. A complex network analysis of such a model offers holistic insight about the performance and risks in such a network. We review the performance and risks of networks with the help of studies that have been done on some of the airport networks. We present various network parameters those could be potentially used as a measure of performance and risks on airport networks. We will also see how various risks, such as break down of airports, spread of diseases across the airport network could be assessed based on the network parameters. Further we review how these insights could possibly be used to 
 shape more efficient and safer airport networks.

\section{\label{intro}Introduction}
Air transportation has become an important component of transportation across the world for long-distance as well as short-distance travel. 
Air transportation has enormous impact on the national and international economies.
Airport networks form the crucial backbone of the air transportation infrastructure. Hence study of airport networks for their performance and the risks posing them, is quite imperative. Airport networks could be classified as complex systems by virtue of their topological as well as their dynamical complexity. Lately there has been growing interest in studying a variety of systems from complex systems viewpoint~\cite{reka:thesis,dorogovtsev:book}. Airports networks too are one of the interesting complex systems which are studied at various scales for various reasons~\cite{air-WAN-PNAS,ani,air-china,air-italy,air-italy02,air-disease,air-WAN,wan-epidemics,air-WAN-model}. According to network dogma, airport network is represented as a graph comprised of `n' nodes (vertices; airports) and `e' links (edges; air-connectivities). Thus represented, airport network looks like a graph (network) whose properties could be computed using graph theoretical formalism. Airport network could
  further be represented as weighted network by considering the (say) number of flights plying on a route as the `weight' of that particular link. Various network parameters give an idea of the performance of the network as well as risks involved in the functioning of the network. 
In this paper, we will discuss various parameters that could be used as a measure of performance and risks on airport networks and discuss how possibly we could construct future airport networks.

\section{\label{performance}Performance of Airport Network}
Airport network is a complex entity by virtue of its topology and traffic dynamics over it. It is a task to define what one means by the performance of the airport network. One way to define performance would be to consider efficient functioning of the network as a whole, while the other could be to consider the ease with which the passengers can travel across the network. Many of the network parameters express efficiency and performance of the network. Following is a list of parameters and features that could serve as a measure of performance of airport network.

\subsubsection*{Characteristic Path Length ($L$)}
Characteristic path length ($L$) is defined as,
\begin{equation}
L = \frac{1}{{N(N-1)}} {\sum_{\substack{i,j=1\\i \ne j}}^N L_{ij}}, 
\label{eq:avg_path_length}
\end{equation}
where, $N$ is total number of nodes in the network, and $L_{ij}$ is the shortest path length between nodes $i$ and $j$.\
Clearly characteristic path length is an average of the shortest path lengths between all possible pairs of nodes. The smaller the $L$, the more compact and reachable the network is. Thus $L$ could be used as an indicator of the performance of the airport network, the performance of the network being inversely proportional to the $L$.

Among the Airport Network of India (ANI) ($L=2.2593$), Airport Network of China (ANC) ($L=2.067$), World-wide Airport Network (WAN) ($L=4.37$), and Italian Airport Network (IAN) ($L=1.98$) the IAN turns out to be most efficient. But it should also be kept in mind that WAN is a much larger network with $3880$ airports whereas IAN has only $42$ airports in it.

\subsubsection*{Clustering Coefficient ($C$)}
Clustering coefficient ($C$) is defined as,
\begin{equation}
C =\frac{1}{N} {\sum_{i=1}^N C_{i}},
\label{eq:avg_cc}
\end{equation}
where $N$ is total number of nodes in the network, and $C_i$ is clustering coefficient of node $i$.
The clustering coefficient of a node is defined as the ratio of number of links amongst its neighboring nodes to the maximum number of links they could have had. It essentially enumerates the probability that two nodes are connected to each other given that they are already (independently) both connected to a common node. 

Clustering coefficient could be used as a measure of performance of airport network. It will represent, on an average, what is the fraction of closed triangles in the airport network. The higher the clustering coefficient the better accessible is the network and hence better is its performance. In an ideal condition the clustering coefficient would be $1$ and hence every airport would be connected to every other airport by a direct air-link. Note that the clustering coefficient in a random network (random connectivities) of the same size and average degree would be significantly smaller, than found in the real-world networks~\cite{ani,air-china}.

By this criterion, among ANI ($C=0.6574$), ANC ($C=0.733$), and IAN ($C=0.10$), the ANC turns out to be the best in terms of reachability as defined by clustering coefficient. Interestingly, IAN has clustering that is comparable to a random model.  

\subsubsection*{Small-World Nature}
High Clustering Coefficient ($C$) along with small Characteristic Path Length ($L$) are two indicators of the small-world nature of the network. But, incidentally all airport networks studied so far have been observed to be small-world networks~\cite{air-WAN-PNAS,ani,air-china,air-italy}, indicating that this network feature is not good enough to adjudge the performance of the network. Perhaps $C$ and $L$ independently are better measures of performance of the network as discussed above.

\subsubsection*{Closeness ($L_i$)}
Closeness ($L_i$) is defined as the average of ($N-1$) shortest paths between node `$i$' and the rest of the nodes. While `Characteristic path length' gives a gross average of shortest paths over the whole network, `Closeness' specifically gives the average of shortest paths that are connected to node `$i$'. Hence `Closeness' is a better measure of connectivity of `a node' to the rest of the network. A plot of `$L_i$ x $i$' gives a complete picture of local connectivity across the whole network. The lower the Closeness of a node, the better is the network connectivity to and from that node, and hence better is the node's performance.

\subsubsection*{``Shortest Path Length'' Plot}
Another way of visualizing the performance of the airport network is to plot Frequency of the Flight-routes versus the Shortest Path Length. So this plot will indicate how many flight-routes exist in the network for a given shortest path length. Ideally, in a well-performing network such a plot should be populated on the low shortest path length side and it should have a nonexistent or a thin tail.

\begin{figure}
\begin{center}
\includegraphics{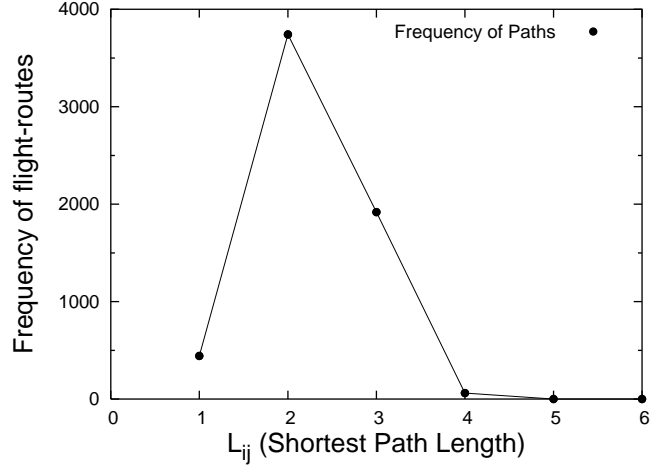}
\end{center}
\caption{\label{fig:shortest_path_dist}Shortest path distribution in Airport Network of India (ANI). The ANI comprises of 
$79$ airports and $449$ one-way flight routes. The network is put together by $11$ airlines, largely domestic and a few 
international.}
\end{figure}

As shown in the Figure~\ref{fig:shortest_path_dist}, the ANI has a peak at shortest path length 2 and has a thin tail between 4 and 5.

\section{\label{risks}Risks on Airport Network}
An airport network could be subjected to variety of risks, varying from traffic congestion, airport shutdown to terrorist attack. In this paper, we will consider risks that involves connectivity, connectivity pattern, breakdown of airports due to natural or human causes, and spread of information (diseases) over the airport network. Earlier, forecast and control of epidemics has been attempted in a world connected by global airport network~\cite{air-disease,wan-epidemics}.

\subsubsection*{Betweenness ($B_k$)}
Betweenness ($B_k$)~\cite{betweenness} of a node `k' is defined as the ratio of number of shortest paths passing through `k' to the total number shortest paths in the network. Essentially Betweenness is a parameter that enumerates the importance of a node in terms of it being central to the traffic in the network. Given that it also represents the importance of the airport (node) to the entire traffic dynamics and hence the risk posed by possible malfunctioning of the airport. This indicates at the list of possible airports that need to be taken special care of, to keep the traffic flow in regulation.
It has been shown for the Italian Airport Network that the betweenness follows a Double Pareto Law~\cite{air-italy}.

\subsubsection*{Coefficient of Assortatvity ($r$)} 
A network is said to show assortative mixing or assortative, if the high-degree nodes in the network tend to be connected with other high-degree nodes, and `disassortative' when the high-degree nodes tend to connect to low-degree nodes~\cite{r:newman}. Clearly this parameter enumerates degree-degree connectivity. It is defined such that it is between -1 and +1. It lies between 0 and +1, if the high degree, high degree connections dominate. In case high degree, poor degree connections are dominant, the coefficient of assortativity lies between 0 and -1.

From computational simulations, it is observed that assortative networks percolate easily~\cite{r:newman}. At the same time the subset of the network to which the percolation is restricted to is `smaller' in the case of assortative network, as opposed to that of disassortative network. This has implication to the spread of diseases over the network. A disease would spread faster on an assortative network, while the set of airports that would form reservoir of the disease would be smaller. Note that WAN (especially weighted WAN) has been shown to be having assortative nature while a regional airport network such as ANI has been shown to be having disassortative nature.

Assortativity also has bearing on resilience. It has been found that assortative networks are resilient to simple targeted attacks~\cite{r:newman}.  In assortatvive networks, removing high-degree nodes is a relatively inefficient strategy for destroying network connectivity. This implies that to avoid destruction of network connectivity due to node (airport) malfunctioning as a result of natural disaster or because of human cause, it is better to have the network with assortative degree mixing. 

\section{\label{future}Design of Future Airports}
Understanding of network parameters that relate to performance and risk on the airport may not be simply be an academic issue, restricted to theoretical studies. The implications coming out of theoretical and computational studies could well be used for implementing into real-life airport networks.  

Section~\ref{performance} offers us $L$, $C$ and $L_i$ as possible measures of network performance. Incidentally, in the real-life airport networks all networks have a low $L$ and significantly high $C$ (except for IAN) compared to the random network model. It is not clear whether decreasing $L$ further or stretching $C$ to it's maximum possible value would improve the performance of the network in terms of making it easy to travel across the network. To improve the connectivity of an airport ,`i', with rest of network, simulations could be done to figure out the network topology for best possible value of $L_i$. At any point of time one may have possible alterations to the topology of the network of which only a few may be better in terms of having improved network performance. 

The degree, number of flight links that a airport has, of an airport is not necessarily linked to its traffic centrality as defined by betweenness ($B_k$). One could use betweenness as a parameter to decide traffic dynamics-wise important airports and provide some appropriate facilities there.

Keeping in view the above points mentioned in Section~\ref{risks} regarding the risks, if one engineers the network to be disassortative to avoid the fast spread of diseases on the airport network, one is risking disruption of the network in case of airport(s) malfunctioning. And vice versa. This puts us in a dilemma as to how to engineer the degree correlations of future airport networks. Perhaps the unweighted network could be engineered as an assortative network, so that any possible natural or man-made disaster could not easily disrupt it. While at the same time, the weighted network could be designed as a disassortative network, so that diseases could not spread (percolate) fast on this network.  

\section{\label{conclusion}Conclusion}
Air transportation, and thereby, airport networks are increasingly becoming important for transportation across the world. Hence study of airport networks for their performance and risks is of crucial for maintenance and engineering of the airport networks.

By virtue of its nature airport network is amenable for modeling using complex network paradigm. It has discrete elements (airports) that are connected by links (air connectivity). Various network parameters and features could be used as a measure of performance and risks in the airport network. Not every parameter could be useful for this purpose as not every parameter may enumerate the performance or risk on the network. 

Characteristic Path Length ($L$) and Clustering Coefficient ($C$) very well enumerate the performance of the airport network. $L$ is inversely while $C$ is directly proportional to the performance of the network. Incidentally, small-world nature is not necessarily a good indicator of an efficient network. While $L$ and $C$ offer a global view of performance of network, Closeness gives a local view of performance of the network.

The results on spread (percolation) and resilience of the networks suggests that design of future airports is not a straightforward task. While tweaking a parameter might improve a certain feature, it might as well impair some other network feature. One may have to deal with tweaking network parameters of weighted and unweighted airport network simultaneously to achieve the desired result. Importantly tweaking a parameter may be easy or difficult job depending on the task. For example, as an engineer it may be relatively easier to create an assortative unweighted network by introducing some flight routes and rerouting a few. But creating a disassortative weighted network may be a tough task as it involves changing the number of flights which are solely governed by passengers' demand. 


\end{document}